# Intrinsic current drive by electromagnetic electron temperature gradient turbulence in tokamak plasmas

Wen He[1], Lu Wang*[1], Shuitao Peng[1], Weixin Guo[1], and Ge Zhuang[2]

[1]International Joint Research Laboratory of Magnetic Confinement Fusion and Plasma Physics, State Key Laboratory of Advanced Electromagnetic Engineering and Technology, School of Electrical and Electronic Engineering, Huazhong University of Science and Technology, Wuhan, 430074, China

[2]University of Science and Technology of China, No.96, JinZhai Road Baohe District, Hefei, Anhui, 230026, China.

*Author to whom any correspondence should be addressed, E-mail: luwang@hust.edu.cn

**Abstract:** The mean parallel current density evolution equation is presented using electromagnetic (EM) gyrokinetic equation. There exist two types of intrinsic current driving mechanisms resulted from EM electron temperature gradient (ETG) turbulence. The first type is the divergence of residual turbulent flux including a residual stress-like term and a kinetic stress-like term. The second type is named as residual turbulent source, which is driven by the correlation between density and parallel electric field fluctuations. The intrinsic current density driven by the residual turbulent source is negligible as compared to that driven by the residual turbulent flux. The ratio of intrinsic current density driven by EM ETG turbulence to the background bootstrap current density is estimated. The local intrinsic current density driven by the residual turbulent flux for mesoscale variation of turbulent flux can reach about 80% of the bootstrap current density in the core region of ITER standard scenario, but there is no net intrinsic current on a global scale. Based on this, the local intrinsic current driven by EM micro-turbulence and its effects on local modification of the profile of safety factor may be needed to be carefully taken into account in the future device with high $\beta_e$ which is the ratio between electron pressure to the magnetic pressure.

Keywords: intrinsic current, ETG turbulence, residual turbulent flux, residual turbulent source, electromagnetic effects

# 1. Introduction

Current density profile is of great importance for tokamak plasmas since it can affect both confinement time and a variety of magnetohydrodynamic (MHD) instabilities, such as the kink mode, and the tearing mode. The current density profile can be strongly affected by the current driving mechanism. Up to now, various current driving mechanisms have been proposed, such as inductive current drive, neutral-beam-injection (NBI) current drive [1], lower hybrid drift instability current drive [2], electron cyclotron current drive [3], the poloidally asymmetric fueling current drive [4], and so on. One of the particular efficient and economical ways is the bootstrap (BS) current driven by the pressure gradient which has been predicted by the neoclassical theory [5]. It is demonstrated to be consistent with the predictions of the neoclassical theory in some experiments [6-8]. Nonetheless, deviations from the neoclassical theory are also observed in some cases [9, 10]. Inspired by the turbulent-driven intrinsic rotation, naturally, intrinsic current driven by turbulence can be taken into account similarly.

In fact, current driven by drift waves in a slab geometry has been proposed many years ago [11]. It showed that a residual current flux could be caused by the electrostatic (ES) fluctuations in the presence of $k_\parallel$ (parallel wave number of drift wave) symmetry breaking. A more systematic model for current driven by turbulence has been done by Garbet [12], where two kinds of current source from the gyrokinetic equation were proposed. Both of them require $k_\parallel$ symmetry breaking and may contribute about 10% of the BS current density in the edge region of tokamak plasmas. Similarly, an extended Ohm's law modified by magnetic turbulence has also been studied by using the self-consistent action-angel transport theory [13]. In this model, turbulence results in three different kinds of resistances in Ohm's law. The modification of Ohm's law by two kinds of current source from electromagnetic (EM) turbulence has been done [14]. One is the momentum source and the other one is radial flux of the parallel current driven by magnetic flutter. Though the momentum source term can bring a little change of the local current density, it contributes a nonzero total current. In contrast, the flux term will significantly change the local current density without a net total current. It is worth mentioning that a novel anomalous BS current can be caused by turbulent scattering [15-17], which is thought to serve as effective collision. The ES trapped electron mode (TEM) turbulence and ion temperature gradient (ITG) turbulence were considered in [17].

Although lacking of experimental evidence, in a slab geometry, the current driven by ES electron temperature gradient (ETG) turbulence has been calculated using a fluid model [18], in which the ratio of current density driven by ETG turbulence to the BS current density can be 10%. Recently, a gyrokinetic simulation result also reports that ES ETG can drive current [19], and it shows that ETG turbulence driven current density reaches about 20% of the BS current density, thus significantly changes the safety factor. On the other hand, experimental evidence found that EM effects may be considerable for intrinsic rotation [20, 21], and theoretical work also addressed significant EM effects on intrinsic rotation drive in the H-mode pedestal plasmas [22]. Thus, investigation of intrinsic current driven by EM ETG turbulence in toroidal tokamak plasmas is meaningful.

In this work, the mean parallel current density evolution equation is derived using the EM gyrokinetic equation. It is shown that there are two types of intrinsic current driving mechanisms. Similar to the turbulent-driven intrinsic rotation, the first type is the divergence of residual turbulent flux, which comes from a residual stress-like term and a kinetic stress-like term. The second type is named as residual turbulent source, which is driven by the correlation between density and parallel electric field fluctuations. The quasilinear estimation of the divergence of the residual turbulent flux and the residual turbulent source of current density driven by EM ETG turbulence are presented. Both of the driving mechanisms need $k_\parallel$ symmetry breaking. By taking the typical core parameters in International Thermonuclear Experimental Reactor (ITER) standard scenario, we compare the intrinsic current density driven by EM ETG turbulence to the background BS current density. The results show that the EM effects could be important for relatively high $\beta_e$ which is the ratio between electron pressure to the magnetic pressure. Although the contribution to the total net current is neglected, the local modification of current density profile due to EM ETG turbulence could play a significant role in future device.

The remainder of this paper is organized as follows. In section 2, the derivation of the mean parallel current density evolution equation is presented. The quasi-linear estimation of intrinsic current density driven by EM ETG turbulence and the comparison between the intrinsic current density and the BS current density are also presented. Finally, conclusions and some discussions are given in section 3.

## 2. Quasilinear estimate for intrinsic parallel current drive

We start from the nonlinear EM gyrokinetic equation [23],

$$\frac{\partial (F_e B_\parallel^*)}{\partial t} + \nabla \cdot \left(\frac{d\mathbf{R}}{dt} F_e B_\parallel^*\right) + \frac{\partial}{\partial v_\parallel}\left(\frac{dv_\parallel}{dt} F_e B_\parallel^*\right) = 0, \quad (1)$$

with the electron gyro-center equations of motion in the symplectic formulation, i.e., $v_\parallel$ representation

$$\frac{d\mathbf{R}}{dt} = v_\parallel \frac{\mathbf{B}^*}{B_\parallel^*} + \frac{c\hat{\mathbf{b}}}{-eB_\parallel^*} \times [(-e\nabla\delta\phi) + \mu\nabla B], \quad (2)$$

and

$$\frac{dv_\parallel}{dt} = -\frac{\mathbf{B}^*}{m_e B_\parallel^*} \cdot (-e\nabla\delta\phi + \mu\nabla B) + \frac{e}{m_e c}\frac{\partial \delta A_\parallel}{\partial t}. \quad (3)$$

Here, $F_e = F_e(\mathbf{R}, \mu, v_\parallel, t)$ is the electron gyro-center distribution function where $\mu$ is the magnetic moment of electron, $\hat{\mathbf{b}} = \mathbf{B}/B$ is the unit vector of equilibrium magnetic line, $B_\parallel^* = \hat{\mathbf{b}} \cdot \mathbf{B}^*$ is the Jacobian of the transformation from the particle phase space to the gyro-center phase space with $\mathbf{B}^* = \mathbf{B} + \delta\mathbf{B}_\perp + \frac{cm_e}{-e}v_\parallel \nabla \times \hat{\mathbf{b}}$, $c$ is the speed of light, $m_e$ is the electron mass, $e$ is the elementary charge. $\delta\phi$ is the ES potential fluctuation. $\delta\mathbf{B}_\perp \approx -\hat{\mathbf{b}} \times \nabla \delta A_\parallel$ is the perturbed magnetic field. In this paper, the index ∥ refers to the components parallel to the equilibrium magnetic field, and the index ⊥ refers to the components perpendicular to the equilibrium magnetic field. Only the shear component of magnetic perturbation, i.e, $\delta A_\parallel$, is considered, and $\delta B_\parallel$ is not included in this work. This simplified model was also widely adopted in previous works on investigation of EM ETG [24-26]. However, for spherical tokamak where $\beta$ can reach 10% or even higher, neglecting $\delta B_\parallel$ will lead to an under estimation of the growth rates of ETG as well as ITG [27].

Multiplying Eq. (1) by $-\frac{2\pi}{m_e}ev_\parallel$ and integrating in the velocity space, the evolution equation of parallel current density can be obtained,

$$\frac{\partial J_\parallel}{\partial t} + \nabla \cdot \left(3v_{d\kappa}J_\parallel + v_{d\nabla}J_\parallel + \delta v_{E\times B}J_\parallel + 2U_0 J_\parallel \widehat{\boldsymbol{b}}^* - eU_0^2 n_e \widehat{\boldsymbol{b}}^*\right)$$
$$+\left(\frac{e^2 \widehat{\boldsymbol{b}}^* \cdot \nabla \delta\phi}{m_e} + \frac{e^2}{m_e}\frac{\partial \delta A_\parallel}{\partial t}\right)n_e + \left(-\frac{e}{m_e}\widehat{\boldsymbol{b}}^* \cdot \nabla P_\parallel\right)$$
$$+\frac{e}{m_e}\widehat{\boldsymbol{b}}^* \cdot \frac{\nabla B}{B}(P_\parallel - P_\perp) + J_\parallel \boldsymbol{v}_{d\kappa}\cdot\left(\frac{-e\nabla\delta\phi}{T_e} + \frac{\nabla B}{B}\right) = 0. \qquad (4)$$

Here, we assume the current is mainly carried by electron. $J_\parallel = -e\int F_e v_\parallel d^3v = -en_e U_\parallel$ is the parallel current density with $n_e$ being the density of electron and $U_\parallel$ being the parallel speed of the electron fluid. $P_\parallel = m_e \int F_e (v_\parallel - U_\parallel)^2 d^3v$ is the parallel electron pressure. $P_\perp = \int F_e \mu B d^3v$ is the perpendicular electron pressure. $\widehat{\boldsymbol{b}}^* = \widehat{\boldsymbol{b}} + \delta \boldsymbol{b}_\perp$ with $\delta \boldsymbol{b}_\perp = \frac{\delta \boldsymbol{B}_\perp}{B}$. $\delta \boldsymbol{v}_{E\times B} = \frac{c\widehat{\boldsymbol{b}}\times\nabla\delta\phi}{B}$ is the fluctuating $\boldsymbol{E}\times\boldsymbol{B}$ drift velocity. $\boldsymbol{v}_{d\nabla} = \frac{cT_\perp}{-eB^2}\widehat{\boldsymbol{b}}\times\nabla B$ is the magnetic gradient drift velocity and $\boldsymbol{v}_{d\kappa} = \frac{cT_\parallel}{-eB}\widehat{\boldsymbol{b}}\times(\widehat{\boldsymbol{b}}\cdot\nabla)\widehat{\boldsymbol{b}}$ is the magnetic curvature drift velocity. The magnetic drift velocity will be neglected in the mean parallel current density equation, since they are higher order terms $O(\omega_{de}/\omega)$ as compared to $\boldsymbol{E}\times\boldsymbol{B}$ drift velocity [28]. For the same reason, all the terms in Eq. (4) related to magnetic gradient will be neglected. The terms related to the equilibrium electron fluid velocity $U_0$ will also be neglected because of $U_0^2 \ll v_{the}^2$, where $v_{the} = \sqrt{\frac{T_e}{m_e}}$ is the electron thermal velocity with $T_e$ being the electron temperature. Taking flux-average of Eq. (4), the evolution of mean parallel current density can be obtained,

$$\frac{\partial \langle J_\parallel \rangle}{\partial t} + \nabla \cdot \langle \delta \boldsymbol{v}_{E\times B,r} \delta J_\parallel \rangle - \nabla \cdot \langle \frac{e}{m_e}\delta \boldsymbol{b}_r \delta P_\parallel \rangle$$
$$= -\frac{e^2}{cm_e}\langle \frac{\partial \delta A_\parallel}{\partial t}\delta n_e \rangle - \frac{e^2}{m_e}\langle \widehat{\boldsymbol{b}}\cdot\nabla\delta\phi \delta n_e \rangle. \qquad (5)$$

Here, $\delta J_\parallel = -e\int \delta f_e v_\parallel d^3v$ is the perturbation of parallel current density with $\delta f_e$

being the perturbed electron distribution function. $\delta P_\parallel = m_e \int \delta f_e (v_\parallel - U_\parallel)^2 d^3 v$ is the perturbation of the parallel electron pressure. The two terms under the divergence on the left hand side (LHS) of Eq. (5) are turbulent flux of current density. Similar with the intrinsic rotation drive in [22], the first one is the Reynolds stress-like term, and the second one is the kinetic stress-like term denoting dynamo effects. The terms on the right hand side (RHS) are turbulent source terms, which is analogous to the momentum source for ion [29]. The source terms are driven by the correlation between density and parallel electric field fluctuations including inductive electric field. Eq. (5) seems to be similar with Eq. (3) derived from fluid model in [18]. However, the ES limit was adopted in quasilinear calculation of intrinsic current drive in [18]. Both the EM effects and toroidal effects are kept in our quasilinear calculation which will be shown later. Then, Eq. (5) can be rewritten as

$$\frac{\partial \langle J_\parallel \rangle}{\partial t} + \nabla \cdot \Gamma_r = S, \qquad (6)$$

where $\Gamma_r$ is the turbulent flux of current density, and $S$ represents the turbulent source. The turbulent flux can be usually divided into diffusive term, convective term and residual term, i.e.,

$$\Gamma_r = -\chi_\parallel \frac{\partial}{\partial r} \langle J_\parallel \rangle + V_c \langle J_\parallel \rangle + \Gamma_r^{res}, \qquad (7)$$

where $\chi_\parallel$ is the diffusion coefficient of the parallel current density, $V_c$ is the convective velocity of the parallel current density and $\Gamma_r^{res}$ is the residual turbulent flux. The residual turbulent flux is independent of the parallel current density or its gradient, so it can provide intrinsic current drive. This is analogous to the intrinsic rotation drive by residual stress [29, 30]. The turbulent source term can be divided into residual turbulent source and non-residual turbulent source terms, similarly. The residual turbulent source is also independent of the parallel current density or its gradient, and so can provide intrinsic current drive, too. This is also analogous to the intrinsic rotation drive by momentum source [29].

Next, we make quasi-linear estimation for the turbulent flux and the residual turbulent source. Therefore, the linear calculations of $\delta n_e$, $\delta J_\parallel$, and $\delta P_\parallel$ are required. The electron density fluctuation can be obtained using the quasi-neutrality condition $\delta n_e = \delta n_i$ and adiabatic ion approximation $\delta n_i = -\frac{e\delta\phi}{T_i}n_0$ for EM ETG mode. The adiabatic ion model is often employed in studying ETG turbulence [25, 26]. Especially, the validity of adiabatic ion model used for ETG turbulence at low magnetic shear was demonstrated via parameter scan in magnetic shear in [31]. The $i\delta$ model is used in [18, 24], which is analogous to the $i\delta$ model for electrons used in ITG turbulence. Assuming the equilibrium electron distribution function to be shifted Maxwellion $F_{e0} = n_0 \left(\frac{m_e}{2\pi T_e}\right)^{3/2} \exp\left(-\frac{m_e(v_\parallel - U_0)^2}{2T_e} - \frac{\mu B}{T_e}\right)$ where $n_0$ is the equilibrium electron density, the linearized EM perturbed electron distribution function in Fourier space can be written as

$$\delta f_{ek} = i \frac{\left\{-\omega_{*e}\left[1 + \eta_e(x_\parallel^2 + x_\perp^2 - 3)\right] + \omega_{de}\left(x_\parallel \hat{v}_\parallel + \frac{1}{2}x_\perp^2\right) + k_\parallel v_{the} x_\parallel - x_\parallel \omega_A\right\} \delta\hat{\phi}_k F_{e0}}{-i\left[\omega_k - \omega_{de}\left(\hat{v}_\parallel^2 + \frac{1}{2}x_\perp^2\right) - k_\parallel v_\parallel\right]}$$

$$- i \frac{\left\{-\omega_{*e}\hat{v}_\parallel\left[1 + \frac{1}{2}\eta_e(x_\parallel^2 + x_\perp^2 - 3)\right] + \omega_k x_\parallel + \frac{x_\perp^2}{2}\hat{U}_0 \omega_{de} - \omega_A x_\parallel \hat{v}_\parallel\right\} \delta\hat{A}_{\parallel,k} F_{e0}}{-i\left[\omega_k - \omega_{de}\left(\hat{v}_\parallel^2 + \frac{1}{2}x_\perp^2\right) - k_\parallel v_\parallel\right]}.$$

(8)

Here, $\rho_e = \frac{v_{the}}{\Omega_e}$ is the gyro-radius of electron, $\Omega_e = \frac{eB}{cm_e}$ is the Larmor frequency of electron. $\delta\hat{A}_{\parallel,k} = \frac{\delta A_{\parallel,k}}{\rho_e B}$, $\delta\hat{\phi}_k = \frac{e\delta\phi_k}{T_e}$, $\hat{v}_\parallel = \frac{v_\parallel}{v_{the}}$, $\hat{U}_0 = \frac{U_0}{v_{the}}$, $x_\parallel = \frac{v_\parallel - U_0}{v_{the}}$ and $x_\perp = \sqrt{\frac{2\mu B}{T_e}}$. $\omega_{*e} = \frac{cT_e}{-eB}\hat{b} \times \nabla \ln n_0 \cdot \boldsymbol{k}$ is the diamagnetic drift frequency of electron with $\boldsymbol{k}$ being wave vector, $\omega_A = \frac{cT_e \hat{b} \times \nabla U_0}{-eB v_{the}} \cdot \boldsymbol{k}$ is the frequency related to the gradient of background electron fluid velocity, $\omega_{de} = \frac{cT_e}{-eB^2}\hat{b} \times \nabla B \cdot \boldsymbol{k} \simeq \frac{cT_e}{-eB^2}\hat{b} \times (\hat{b} \cdot \nabla \hat{b}) \cdot \boldsymbol{k} \simeq \frac{k_\theta \rho_e v_{the}}{R}$ is the drift frequency due to magnetic gradient or curvature. This is equivalent to taking the position at outboard midplane. It should be pointed out that the complex poloidal angle dependence of magnetic drift frequency is neglected for simplicity. Here, $\hat{b} \times (\hat{b} \cdot \nabla \hat{b}) \simeq \hat{b} \times \nabla \ln B$ is used, and is justified even for high beta case as long as $\beta \ll 1$. $\eta_e = L_n/L_{Te}$ with $L_{Te} = -(\nabla \ln T_e)^{-1}$ and $L_n =$ $-(\nabla \ln n_e)^{-1}$ being the electron temperature gradient length scale and electron density gradient length scale, respectively. On the RHS of Eq. (8) the first term comes from ES contribution, and the second term is the contribution from EM effects. If the condition $\omega_{de} > k_\parallel v_{the}$ is satisfied, we can neglect $k_\parallel v_\parallel$ in the denominator of Eq. (8) for simplicity. The magnetic drift resonance for current drive is stronger than the transit resonance. This is possible for relatively short wavelength ETG turbulence and weak magnetic shear, which will be discussed later. Here, only the magnetic drift resonance is considered. Of course, it would be more accurate to simultaneously consider both the transit frequency and the magnetic drift frequency in the denominator. However, it is too difficult for analytical treatment in that way. The inverse electron propagator can be written as

$$\left\{-i\left[\omega_k - \omega_{de}\left(\hat{v}_\parallel^2 + \frac{1}{2}x_\perp^2\right)\right]\right\}^{-1} \simeq \pi\delta\left[\omega_k - \omega_{de}\left(\hat{v}_\parallel^2 + \frac{1}{2}x_\perp^2\right)\right] + \frac{i}{\omega_k}. \quad (9)$$

The lowest order of non-resonant part will be kept.

Taking the first order moment of Eq. (8), i.e., $\delta J_\parallel = -e\int \delta f_e v_\parallel d^3v$ we can obtain the perturbed parallel current density

$$\delta J_{\parallel k} = eU_0\tau n_0 \delta\hat{\phi}_k + \frac{\hat{U}_0\omega_{de} - \omega_A + k_\parallel v_{the}}{\omega_k} en_0 v_{the}\delta\hat{\phi}_k$$

$$+ \frac{[\omega_{*e}(1+\eta_e) - \omega_k + \hat{U}_0\omega_A]}{\omega_k} en_0 v_{the}\delta\hat{A}_{\parallel,k}$$

$$- i\frac{3\sqrt{3\pi}}{8}\frac{\hat{U}_0\omega_{de} - \omega_A + k_\parallel v_{the}}{|\omega_{de}|}\left(\frac{\omega_k}{\omega_{de}}\right)^{3/2}\exp\left(-\frac{3}{4}\frac{\omega_k}{\omega_{de}}\right)en_0 v_{the}\delta\hat{\phi}_k$$

$$+ i\frac{3\sqrt{3\pi}}{8}\left(\frac{\omega_k}{\omega_{de}}\right)^{3/2}\left\{-\omega_{*e}\left[1+\eta_e\left(\frac{3}{4}\frac{\omega_k}{\omega_{de}} - \frac{3}{2}\right)\right] + \omega_k\right.$$

$$\left. - \omega_A\hat{U}_0\right\}\frac{1}{|\omega_{de}|}\exp\left(-\frac{3}{4}\frac{\omega_k}{\omega_{de}}\right)en_0 v_{the}\delta\hat{A}_{\parallel,k},$$

(10)

where $\tau = \frac{T_e}{T_i}$ is the ratio of electron temperature to ion temperature. Here, the assumption $2\hat{v}_\parallel^2 = x_\perp^2$ has been used for calculation of the resonant part. The first two terms in Eq. (10) result from non-resonant ES contribution, the third term comes from non-resonant EM effects, and the last two terms are caused by resonant effects.

Taking the second order moment, i.e., $\delta P_{\parallel} = m_e \int \delta f_e (v_{\parallel} - U_{\parallel})^2 d^3v \simeq m_e \int \delta f_e (v_{\parallel} - U_0)^2 d^3v$, we can obtain the perturbed electron pressure,

$$\delta P_{\parallel,k} = \frac{1}{\omega_k}[\omega_{*e}(1+\eta_e) - 4\omega_{de}]n_0 T_e \delta\hat{\phi}_k$$
$$- \frac{\omega_{*e}(1+\eta_e)\hat{U}_0 - \omega_{de}\hat{U}_0 + 3\omega_A}{\omega_k} n_0 T_e \delta\hat{A}_{\parallel,k}$$
$$- i\frac{3\sqrt{3\pi}}{8}\frac{1}{|\omega_{de}|}\left(\frac{\omega_k}{\omega_{de}}\right)^{3/2} \exp\left(-\frac{3}{4}\frac{\omega_k}{\omega_{de}}\right)\left\{\omega_{*e}\left[1+\eta_e\left(\frac{3}{4}\frac{\omega_k}{\omega_{de}} - \frac{3}{2}\right)\right] - \omega_k\right\}n_0 T_e \delta\hat{\phi}_k$$
$$+ i\frac{3\sqrt{3\pi}}{8}\left(\frac{\omega_k}{\omega_{de}}\right)^{3/2}\left\{\omega_{*e}\hat{U}_0\left[1+\eta_e\left(\frac{3}{4}\frac{\omega_k}{\omega_{de}} - \frac{3}{2}\right)\right] - \frac{1}{2}\hat{U}_0\omega_k\right.$$
$$\left. + \frac{1}{2}\frac{\omega_k}{\omega_{de}}\omega_A\right\}\frac{1}{|\omega_{de}|}\exp\left(-\frac{3}{4}\frac{\omega_k}{\omega_{de}}\right)n_0 T_e \delta\hat{A}_{\parallel,k}.$$

(11)

Similar to Eq. (10), the first two terms on the RHS of Eq. (11) are non-resonant parts, and the last two terms are resonant parts.

We take $\omega_k = \omega_{kr} + i\gamma_k$ with $\omega_{kr}$ being the real frequency and $\gamma_k$ being the linear growth rate, respectively. The higher orders of $\gamma_k$ will be neglected because of $|\gamma_k|^2 \ll \omega_{kr}^2$ which justifies the quasi-linear theory. Then, using $\delta v^*_{E\times B,r} = \sum_k ik_\theta \rho_e v_{the} \delta\hat{\phi}_{-k}$, $\omega_A = -k_\theta\rho_e \frac{\partial U_0}{\partial r}$, $\omega_{*e} = k_\theta\rho_e \frac{v_{the}}{L_n}$, $\delta b_{rk} = ik_\theta\rho_e \delta\hat{A}_{\parallel,k}$, we can calculate the turbulent flux terms and the turbulent source terms. The details of the calculation are presented in the Appendix. Here, we just directly write the expressions,

$$\Gamma_r = -\sum_k \frac{\gamma_k}{\omega_{kr}^2} k_\theta^2 \rho_e^2 v_{the}^2 \left(|\delta\hat{\phi}_k|^2 + 3|\delta\hat{A}_{\parallel,k}|^2\right)\frac{\partial\langle J_{\parallel}\rangle}{\partial r}$$
$$-\sum_k \frac{3\sqrt{3\pi}}{8} k_\theta^2\rho_e^2 \frac{v_{the}^2}{|\omega_{de}|}\left(\frac{\omega_{kr}}{\omega_{de}}\right)^{3/2}\exp\left(-\frac{3}{4}\frac{\omega_{kr}}{\omega_{de}}\right)\left(|\delta\hat{\phi}_k|^2 + \frac{1}{2}\frac{\omega_{kr}}{\omega_{de}}|\delta\hat{A}_{\parallel,k}|^2\right)\frac{\partial\langle J_{\parallel}\rangle}{\partial r}$$
$$-\sum_k \frac{\gamma_k}{\omega_{kr}^2} k_\theta\rho_e v_{the}\left[(\omega_{de} + \omega_{*e})|\delta\hat{\phi}_k|^2 + (\omega_{de} + 2\omega_{*e} - \eta_e\omega_{*e})|\delta\hat{A}_{\parallel,k}|^2\right]\langle J_{\parallel}\rangle$$

$$-\sum_k \frac{3\sqrt{3\pi}}{8} k_\theta \rho_e v_{the} \left(\frac{\omega_{kr}}{\omega_{de}}\right)^{3/2} \exp\left(-\frac{3}{4}\frac{\omega_{kr}}{\omega_{de}}\right) \left\{\left(\frac{\omega_{de}}{|\omega_{de}|} + \frac{\omega_{*e}}{|\omega_{de}|}\right)|\delta\hat{\phi}_k|^2\right.$$

$$\left. - \left\{\frac{\omega_{*e}}{|\omega_{de}|}\left[1 + \eta_e\left(\frac{3}{4}\frac{\omega_{kr}}{\omega_{de}} - \frac{3}{2}\right)\right] - \frac{1}{2}\frac{\omega_{kr}}{|\omega_{de}|} - \frac{1}{2}\frac{\omega_{*e}}{|\omega_{de}|}\frac{\omega_{kr}}{\omega_{de}}\right\}|\delta\hat{A}_{\parallel,k}|^2\right\}\langle J_\parallel\rangle$$

$$+ \sum_k \frac{\gamma_k}{\omega_{kr}^2} k_\theta \rho_e e n_0 v_{the}^2 k_\parallel v_{the} |\delta\hat{\phi}_k|^2$$

$$+ \sum_k \left(1 - \frac{4\omega_{de}}{\omega_{kr}}\right) k_\theta \rho_e e n_0 v_{the}^2 \text{Im}\langle\delta\hat{A}_{\parallel,k}\delta\hat{\phi}_{-k}\rangle$$

$$+ \sum_k \frac{\gamma_k[2\omega_{*e}(1+\eta_e) - 4\omega_{de}]}{\omega_{kr}^2} k_\theta \rho_e e n_0 v_{the}^2 \text{Re}\langle\delta\hat{A}_{\parallel,k}\delta\hat{\phi}_{-k}\rangle$$

$$+ \sum_k \frac{3\sqrt{3\pi}}{8} k_\theta \rho_e e n_0 v_{the}^2 \frac{k_\parallel v_{the}}{|\omega_{de}|}\left(\frac{\omega_{kr}}{\omega_{de}}\right)^{3/2} \exp\left(-\frac{3}{4}\frac{\omega_{kr}}{\omega_{de}}\right)|\delta\hat{\phi}_k|^2$$

$$+ \sum_k \frac{6\sqrt{3\pi}}{8} k_\theta \rho_e e n_0 v_{the}^2 \left(\frac{\omega_{kr}}{\omega_{de}}\right)^{3/2} \exp\left(-\frac{3}{4}\frac{\omega_{kr}}{\omega_{de}}\right)\left\{\frac{\omega_{*e}}{|\omega_{de}|}\left[1\right.\right.$$

$$\left.\left. + \eta_e\left(\frac{3}{4}\frac{\omega_{kr}}{\omega_{de}} - \frac{3}{2}\right)\right] - \frac{\omega_{kr}}{|\omega_{de}|}\right\}\text{Re}\langle\delta\hat{A}_{\parallel,k}\delta\hat{\phi}_{-k}\rangle,$$

(12)

and

$$S = \sum_k e n_0 v_{the}\, \tau\big(\omega_{kr}\text{Im}\langle\delta\hat{A}_{\parallel,k}\delta\hat{\phi}_{-k}\rangle + \gamma_k\text{Re}\langle\delta\hat{A}_{\parallel,k}\delta\hat{\phi}_{-k}\rangle\big).$$

(13)

Different from [12], the non-resonant parts are also calculated here. Moreover, we will consider the relation between $\delta\hat{A}_\parallel$ and $\delta\hat{\phi}$ in more detail to investigate the explicit EM effects. In the following, we will focus on the residual turbulent flux and the residual source, which contribute to the intrinsic current drive. Combing the Ampere's law with Eq. (10) and neglecting terms related to $U_0$, the general relation between $\delta\hat{\phi}$ and $\delta\hat{A}_\parallel$ can be obtained,

$$\delta\hat{A}_{\parallel,k} = \frac{D_1(C_1-C_2) - (D_2+D_3)(C_3+C_4) - i[(D_2+D_3)(C_1-C_2) + D_1(C_3+C_4)]}{(C_1-C_2)^2 + (C_3+C_4)^2} \delta\hat{\phi}_k. \quad (14)$$

Here, $C_1, C_2, C_3, C_4$ and $D_1, D_2, D_3$ are all dimensionless and independent of parallel current density or gradient of parallel current density. Eq. (14) is important for explicit estimate of the EM effects. In [12], this relation is not mentioned and the EM effects are not the focus. Besides, the spatial scale of turbulence is about ion gyroradius in [12]. The details of the calculation of Eq. (14) can be found in the Appendix.

In the following, we assume $\omega_{kr} \simeq \omega_{*e}$ which is appropriate for ETG mode. The residual turbulent flux can be written as

$$\Gamma_r^{res} = \sum_k \frac{\gamma_k}{\omega_{kr}^2} k_\theta \rho_e e n_0 v_{the}^2 k_\parallel v_{the} |\delta \hat{\phi}_k|^2$$

$$+ \sum_k \left(1 - \frac{4\omega_{de}}{\omega_{kr}}\right) k_\theta \rho_e e n_0 v_{the}^2 \text{Im}\langle \delta \hat{A}_{\parallel,k} \delta \hat{\phi}_{-k}\rangle$$

$$+ \sum_k \frac{\gamma_k[2\omega_{*e}(1+\eta_e) - 4\omega_{de}]}{\omega_{kr}^2} k_\theta \rho_e e n_0 v_{the}^2 \text{Re}\langle \delta \hat{A}_{\parallel,k} \delta \hat{\phi}_{-k}\rangle$$

$$+ \sum_k \frac{3\sqrt{3\pi}}{8} k_\theta \rho_e e n_0 v_{the}^2 \frac{k_\parallel v_{the}}{|\omega_{de}|} \left(\frac{\omega_{kr}}{\omega_{de}}\right)^{3/2} \exp\left(-\frac{3}{4}\frac{\omega_{kr}}{\omega_{de}}\right) |\delta \hat{\phi}_k|^2$$

$$+ \sum_k \frac{6\sqrt{3\pi}}{8} k_\theta \rho_e e n_0 v_{the}^2 \left(\frac{\omega_{kr}}{\omega_{de}}\right)^{\frac{3}{2}} \exp\left(-\frac{3}{4}\frac{\omega_{kr}}{\omega_{de}}\right) \left\{\frac{\omega_{*e}}{|\omega_{de}|}\left[1\right.\right.$$

$$\left.\left.+ \eta_e\left(\frac{3}{4}\frac{\omega_{kr}}{\omega_{de}} - \frac{3}{2}\right)\right] - \frac{\omega_{kr}}{|\omega_{de}|}\right\} \text{Re}\langle \delta \hat{A}_{\parallel,k} \delta \hat{\phi}_{-k}\rangle.$$

(15)

For turbulent source, $S^{res} = S$ which is also independent of parallel current density or its gradient due to Eq. (14), and so can provide an intrinsic current drive. The first term on the RHS of Eq. (15) represents the non-resonant ES contribution, the second term and the third term represent the non-resonant EM contribution, the forth term represents resonant ES contribution, and the last term represents resonant EM contribution. We will use $\frac{\omega_{de}}{\omega_{*e}} \simeq \frac{L_n}{R}$ later. The terms related to $\frac{L_n}{R}$ come from toroidal effects. Eq. (13) shows that the residual turbulent source is only caused by EM effects, since the turbulent source driven by the correlation between density and ES field fluctuations vanishes for the adiabatic ion response. Both the residual turbulent flux and the residual turbulent source require parallel symmetry breaking. Theoretical works have proposed various symmetry breaking mechanisms, such as $\boldsymbol{E} \times \boldsymbol{B}$ shear [32, 33], charge separation from polarization drift [34], intensity gradient [35], geometrical up-down asymmetries [36], etc. In [19], it is found that the contribution to the ETG turbulence driven current from symmetry breaking induced by the turbulence intensity gradient is more important than that by zonal flow shear from ES gyrokinetic simulations. We also take the symmetry breaking caused by the turbulence intensity gradient in this work. This is reasonable for core region with flat pressure profile where the $\boldsymbol{E} \times \boldsymbol{B}$ shear is not strong. The mean parallel wave number is $\bar{k}_\parallel R \simeq \hat{s} k_\theta \frac{w_k^2}{L_I}$ with $\bar{k}_\parallel = \sum_{\vec{k}} k_\parallel |\delta\hat{\phi}_k|^2 / \sum_{\vec{k}} |\delta\hat{\phi}_k|^2$, $R$ being the major radius, $\hat{s}$ being the magnetic shear, $w_k$ being the mode width and $L_I = \left(\frac{d\ln(I_k)}{dr}\right)^{-1}$ being the turbulence intensity gradient length scale [12, 35] with $I_k = |\delta\hat{\phi}_k|^2$.

The turbulence driven intrinsic current density can be estimated by balancing the negative divergence of residual turbulent flux and residual turbulent source with the collisional friction force $-v_{ei}J_{turb}$. The intrinsic current density driven by the residual turbulent flux can be written as

$$J_{turb}^\Gamma = \mp \frac{\Gamma_r^{res}}{v_{ei}\sqrt{\rho_e L_n}}. \tag{16}$$

Here, the length scale of variation of the residual turbulent flux is taken as mesoscale, i.e., $\sqrt{\rho_e L_n}$. There does not exist experimental evidence for the length scale. The sign of $\mp$ corresponds to positive (negative) gradient of flux. In [18], only the ES part, namely the first term on the RHS of Eq. (15) is considered. While the toroidal effects, EM effects and resonant effects are neglected in [18]. Similarly, intrinsic current density driven by residual turbulent source $S^{res}$ can be calculated

$$J_{turb}^S = \frac{S^{res}}{v_{ei}}. \tag{17}$$

The residual turbulent source is caused by momentum exchange between ions and electrons. To illustrate how important the intrinsic current density driven by EM ETG turbulence as compared to the BS current density, we also estimate the BS current density as following [12],

$$J_{BS} \simeq 5\sqrt{\frac{1}{\varepsilon}} \frac{cqn_0 T_e}{BL_P}, \tag{18}$$

where $L_P$ is the length scale of pressure gradient, and $q$ is the safety factor.

Subsequently, the ratios of the intrinsic current density driven by residual turbulent flux and residual turbulent source to the BS current density can be written as

$$\frac{J_{turb}^{\Gamma}}{J_{BS}} = \mp \frac{\sqrt{\varepsilon}\Gamma_r^{res}}{5\nu_{ei}\sqrt{\rho_e L_n}} \frac{L_P}{eqn_0 v_{the}\rho_e}, \tag{19}$$

and

$$\frac{J_{turb}^{S}}{J_{BS}} = \frac{\sqrt{\varepsilon}S^{res}}{5\nu_{ei}} \frac{L_P}{eqn_0 v_{the}\rho_e}. \tag{20}$$

Then, we take the typical parameters of ITER [37] at $r = 0.5a$, $q = 1$, $R/L_{Te} = 4$, $R/L_n = 1.5$, $\hat{s} = 0.3$, $\varepsilon = 0.16$, $R = 6.2$m, $B = 4.85$T, $\tau = 1$, $n_e = 1.3 \times 10^{20}/\text{m}^3$, $T_e = 12$keV, $\nu_{ei} = 5.8 \times 10^3$Hz, $\rho_e = 5.4 \times 10^{-5}$m, $\beta_e = 2.67\%$. It should be stressed out that we have employed the conventional Gaussian units to derive the formulas. The parameters except temperature in international system of units are given here. However, this does not affect the results. The typical ETG scale [31] is taken as $k_\theta \rho_e \simeq 0.4$, $\gamma_k/\omega_{kr} = 3/10$, $w_k = \frac{1}{k_\theta}$, $\sum_k I_k = 10^{-4}$, $L_I = L_n$. Based on these parameters, we can obtain $\omega_{de}/(k_\parallel v_{the}) \sim k_\theta \rho_e q/\hat{s} \sim \frac{4}{3} > 1$, which is consistent with previous assumption. Of course, it would be more accurate to deal with transit and drift resonance simultaneously. However, we have to adopt simplified model and keep some key physics to be able to derive an analytical expression. Note that, the finite Larmor radius effects are neglected in this work, which may quantitatively modify the results and should be carefully considered in future. Numerical results showed that the linear growth rate of EM ETG increases with $k_\theta \rho_e$ [25]. Then, the ratios can be written as

$$\frac{J_{turb}^{\Gamma}}{J_{BS}} = \mp(17.3\% + 9.9\% + 59.6\% - 3.7\%), \tag{21}$$

$$\frac{J_{turb}^{S}}{J_{BS}} < 1\%. \tag{22}$$

On the RHS of Eq. (21), the first term and the third term represent the non-resonant and resonant ES contributions, respectively. While the second term and the last term represent the non-resonant and resonant EM contributions, respectively. For this case, the intrinsic current density driven by residual turbulent flux can reach up to about 80% of the BS current density, but does not contribute to the total net current. On the other hand, the residual turbulent source can drive a nonzero total current, but the order of magnitude is less than 1% as compared to the BS current. Moreover, a rough estimation for ratio of the ETG turbulence driven current density to BS current density by taking similar parameters to those from [19] is about 10% which is comparable to the ratio in [19]. In this paper, the contribution from magnetic drift resonance is calculated since a kinetic model is used. While in [18], only the non-resonant ES contribution was considered due to a fluid model in pedestal region, and the toroidal effects were not taken into account, either. From Eq. (21), it is shown that the resonant contribution is very important as compared to the non-resonant one. This is because the density profile is flat in the core region, making the characteristic frequency of ETG turbulence comparable to the magnetic drift frequency.

The current density profile may be locally modified on the length scale of $\sqrt{\rho_e L_n}$ (about hundreds of electron gyroradii or several ion gyroradii) by the EM ETG turbulence driven intrinsic current in the core region of tokamak H-mode plasmas. This may lead to the local modification of $q$ profile, and hence affect the MHD behaviors. However, if the length scale of the residual turbulent flux is at macroscale, e.g., $L_n$, the ratio of intrinsic current density driven by turbulent flux to the BS current density will reduce to $1/277$ of that estimated for mesoscale turbulent flux. Then, the ratio becomes less than 1%, and the intrinsic current density driven by ETG turbulence is thus negligible for this case. Therefore, the selection of the length scale of the residual turbulent flux is important and deserved to be verified in experiments. A brief summary of the result is given in Table 1.

**Table 1.** Results of the estimation for intrinsic current density driven by EM ETG turbulence for typical parameters core parameters of ITER standard scenario.

| Ratio of intrinsic current density to BS current density | non-resonant contribution | resonant contribution |
|---|---|---|
| ES contribution | $\mp 17.4\%$ | $\mp 59.6\%$ |
| EM contribution | $\mp(9.9\%)$ | $\pm 3.7\%$ |

## 3. Summary

In this work, an evolution equation of mean parallel current density has been derived using the EM gyrokinetic equations. There exist two intrinsic current driving mechanisms. One is the residual turbulent flux and the other is the residual turbulent source. Both of them can provide intrinsic current drive and need $k_\parallel$ symmetry breaking, and the symmetry breaking caused by the turbulence intensity gradient is taken in the present work.

Although the net intrinsic current driven by the EM ETG turbulence can be neglected, the local current density profile can be significantly affected. Quasi-linear estimations show that while the residual turbulent source may not contribute a lot to the current density as compared to the local BS current density, the residual turbulent flux can locally drive about 80% of the local BS current density by using the core parameters of standard scenario ITER. Therefore, we conclude that in the high beta fusion devices like ITER, the EM ETG turbulence driven intrinsic current density may significantly change the local current density profile, thus may change the $q$ profile in core region. This modification of current density profile may be important for MHD instabilities. Particularly, the local modification of current density profile in the narrow pedestal region might be important for edge localized modes, which will be considered in future.

Until now, there is no direct experimental observations for turbulence driven current. However, we expect that in the future fusion reactor with high beta the EM ETG turbulence driven intrinsic current density could be observed. We should point out that the diffusive coefficient and convective speed of current flux are not calculated in the present work. The diffusion and convection may be important for accurate prediction of current density profile in the pedestal region. Moreover, the modification of current density profile in pedestal could affect the edge localized modes control. Therefore, extending this work from core region to pedestal region and considering the effects of diffusion and convection induced by EM turbulence on current density profile may be investigated in future. In addition, collisionless TEM turbulence is another prominent candidate for the electron heat transport in high temperature plasmas. Now, we are also working on the intrinsic current driven by EM collisionless TEM turbulence.

## Acknowledgments

We thank P. H. Diamond, J. Q. Li and W. X. Wang for useful discussions. This work was supported by the Ministry of Science and Technology of China, under Contract No. 2013GB112002, and the NSFC Grant Nos. 11675059 and 11305071.

## Appendix. Calculations of intrinsic current drive and the relation between $\delta \hat{A}_\parallel$ and $\delta \hat{\phi}$

The turbulent current flux is consist of two terms. The non-resonant Reynolds stress-like term can be calculated using Eq. (10),

$$\langle \delta v_{E\times B,r}^* \delta J_\parallel \rangle^{NR} = -\sum_k \frac{\gamma_k k_\theta^2 \rho_e^2 v_{the}^2}{\omega_{kr}^2} |\delta\hat{\phi}_k|^2 \frac{\partial}{\partial r}\langle J_\parallel \rangle - \sum_k \frac{\gamma_k k_\theta \rho_e v_{the}}{\omega_{kr}^2}(\omega_{de}+\omega_{*e})|\delta\hat{\phi}_k|^2 \langle J_\parallel \rangle$$

$$+ \sum_k \frac{\gamma_k k_\parallel k_\theta \rho_e}{\omega_{kr}^2} en_0 v_{the}^3 |\delta\hat{\phi}_k|^2 - \sum_k \frac{\omega_{*e}(1+\eta_e)k_\theta \rho_e}{\omega_{kr}} en_0 v_{the}^2 \mathrm{Im}\langle \delta\hat{A}_{\parallel,k} \delta\hat{\phi}_{-k}\rangle$$

$$+ \sum_k \frac{\gamma_k \omega_{*e}(1+\eta_e)k_\theta \rho_e}{\omega_{kr}^2} en_0 v_{the}^2 \mathrm{Re}\langle \delta\hat{A}_{\parallel,k} \delta\hat{\phi}_{-k}\rangle + \sum_k en_0 v_{the}^2 k_\theta \rho_e \mathrm{Im}\langle \delta\hat{A}_{\parallel,k} \delta\hat{\phi}_{-k}\rangle \ .$$

(A1)

Similarly, the resonant Reynold stress-like term can be obtained,

$$\langle \delta v_{E\times B,r}^* \delta J_\parallel \rangle^R = -\sum_k \frac{3\sqrt{3\pi}}{8} k_\theta^2 \rho_e^2 \frac{v_{the}^2}{|\omega_{de}|} \left(\frac{\omega_{kr}}{\omega_{de}}\right)^{3/2} \exp\left(-\frac{3}{4}\frac{\omega_{kr}}{\omega_{de}}\right) |\delta\hat{\phi}_k|^2 \frac{\partial \langle J_\parallel \rangle}{\partial r}$$

$$- \sum_k \frac{3\sqrt{3\pi}}{8} k_\theta \rho_e v_{the} \left(\frac{\omega_{kr}}{\omega_{de}}\right)^{3/2} \exp\left(-\frac{3}{4}\frac{\omega_{kr}}{\omega_{de}}\right) \left[\frac{\omega_{de}}{|\omega_{de}|} + \frac{\omega_{*e}}{|\omega_{de}|}\right] |\delta\hat{\phi}_k|^2 \langle J_\parallel \rangle$$

$$+ \sum_k \frac{3\sqrt{3\pi}}{8} k_\theta \rho_e e n_0 v_{the}^2 \frac{k_\parallel v_{the}}{|\omega_{de}|} \left(\frac{\omega_{kr}}{\omega_{de}}\right)^{3/2} \exp\left(-\frac{3}{4}\frac{\omega_{kr}}{\omega_{de}}\right) |\delta\hat{\phi}_k|^2$$

$$+ \sum_k \frac{3\sqrt{3\pi}}{8} k_\theta \rho_e e n_0 v_{the}^2 \left(\frac{\omega_{kr}}{\omega_{de}}\right)^{3/2} \exp\left(-\frac{3}{4}\frac{\omega_{kr}}{\omega_{de}}\right) \left\{\frac{\omega_{*e}}{|\omega_{de}|}\left[1 + \eta_e\left(\frac{3}{4}\frac{\omega_{kr}}{\omega_{de}} - \frac{3}{2}\right)\right] - \frac{\omega_{kr}}{|\omega_{de}|}\right\} \mathrm{Re}\langle \delta\hat{A}_{\parallel,k} \delta\hat{\phi}_{-k}\rangle.$$

(A2)

The non-resonant kinetic stress-like term can be calculated using Eq. (11),

$$\langle \frac{e}{m_e}\delta P_\parallel^* \delta b_r\rangle^{NR} = + \sum_k \frac{\gamma_k k_\theta \rho_e v_{the}}{\omega_{kr}^2}(2\omega_{*e} + \omega_{de} - \eta_e \omega_{*e})|\delta\hat{A}_{\parallel,k}|^2 \langle J_\parallel\rangle$$

$$- \sum_k \frac{[\omega_{*e}(1+\eta_e) - 4\omega_{de}]}{\omega_{kr}} k_\theta \rho_e e n_0 v_{the}^2 \mathrm{Im}\langle \delta\hat{A}_{\parallel,k}\delta\hat{\phi}_{-k}\rangle$$

$$- \sum_k \frac{\gamma_k[\omega_{*e}(1+\eta_e) - 4\omega_{de}]}{\omega_{kr}^2} k_\theta \rho_e e n_0 v_{the}^2 \mathrm{Re}\langle \delta\hat{A}_{\parallel,k}\delta\hat{\phi}_{-k}\rangle.$$

(A3)

Similarly, the resonant kinetic-stress like term can be also obtained,

$$\langle \frac{e}{m_e}\delta P_\parallel^* \delta b_r\rangle^R = \sum_k \frac{3\sqrt{3\pi}}{8} k_\theta^2 \rho_e^2 \frac{v_{the}^2}{|\omega_{de}|}\left(\frac{\omega_{kr}}{\omega_{de}}\right)^{3/2} \exp\left(-\frac{3}{4}\frac{\omega_{kr}}{\omega_{de}}\right) \frac{1}{2}\frac{\omega_{kr}}{\omega_{de}}|\delta\hat{A}_{\parallel,k}|^2 \frac{\partial\langle J_\parallel\rangle}{\partial r}$$

$$- \sum_k \frac{3\sqrt{3\pi}}{8} k_\theta \rho_e v_{the} \left(\frac{\omega_{kr}}{\omega_{de}}\right)^{3/2} \exp\left(-\frac{3}{4}\frac{\omega_{kr}}{\omega_{de}}\right) \left(\left\{\frac{\omega_{*e}}{|\omega_{de}|}\left[1 + \eta_e\left(\frac{3}{4}\frac{\omega_{kr}}{\omega_{de}} - \frac{3}{2}\right)\right]\right.\right.$$

$$\left.\left. - \frac{1}{2}\frac{\omega_{kr}}{|\omega_{de}|} - \frac{1}{2}\frac{\omega_{*e}}{|\omega_{de}|}\frac{\omega_{kr}}{\omega_{de}}\right\}|\delta\hat{A}_{\parallel,k}|^2\right)\langle J_\parallel\rangle$$

$$- \sum_k \frac{3\sqrt{3\pi}}{8} k_\theta \rho_e e n_0 v_{the}^2 \left(\frac{\omega_{kr}}{\omega_{de}}\right)^{3/2} \exp\left(-\frac{3}{4}\frac{\omega_{kr}}{\omega_{de}}\right) \left\{\frac{\omega_{*e}}{|\omega_{de}|}\left[1 + \eta_e\left(\frac{3}{4}\frac{\omega_{kr}}{\omega_{de}} - \frac{3}{2}\right)\right] - \frac{\omega_{kr}}{|\omega_{de}|}\right\} \mathrm{Re}\langle \delta\hat{A}_{\parallel,k}\delta\hat{\phi}_{-k}\rangle$$

(A4)

The turbulent source also includes two components. The turbulent source driven by parallel inductive electric field can be written as

$$\frac{e^2}{cm_e}\langle \frac{\partial \delta A_\parallel}{\partial t}\delta n_e^*\rangle = -\sum_k e n_0 v_{the}\tau[\omega_{rk}\mathrm{Im}\langle \delta\hat{A}_{\parallel,k}\delta\hat{\phi}_{-k}\rangle + \gamma_k \mathrm{Re}\langle \delta\hat{A}_{\parallel,k}\delta\hat{\phi}_{-k}\rangle].$$

(A5)

The ES field driven source is

$$\langle \delta n_e \hat{\boldsymbol{b}} \cdot \nabla \delta \phi \rangle = 0. \quad (A6)$$

This is because the adiabatic ion response was used in this work.

The general relation between $\delta A_\parallel$ and $\delta \phi$ can be obtained through Ampere's law

$$-\nabla^2 \delta A_\parallel = \frac{4\pi}{c} \delta J_\parallel. \quad (A7)$$

Neglecting the $U_0$ related terms in Eq. (10) and putting it into Eq. (A7), after some algebra, we can obtain

$$\delta \hat{A}_{\parallel,k} = \frac{D_1(C_1-C_2)-(D_2+D_3)(C_3+C_4)-i[(D_2+D_3)(C_1-C_2)+D_1(C_3+C_4)]}{(C_1-C_2)^2+(C_3+C_4)^2} \delta \hat{\phi}_k. \quad (A8)$$

Here, $C_1 = \frac{2k_\perp^2 \rho_e^2}{\beta_e}$ with $k_\perp$ being the perpendicular wave number and $\beta_e = \frac{8\pi n_e T_e}{B^2}$, $C_2 = \frac{\omega_{*e}(1+\eta_e)}{\omega_{kr}} - 1$, $C_3 = \frac{\gamma_k \omega_{*e}(1+\eta_e)}{\omega_{kr}^2}$, $C_4 = \frac{3\sqrt{3}\pi}{8} \left\{ \frac{\omega_{*e}}{|\omega_{de}|} \left[ 1 + \eta_e \left( \frac{3}{4} \frac{\omega_{kr}}{\omega_{de}} - \frac{3}{2} \right) \right] - \frac{\omega_{kr}}{|\omega_{de}|} \right\} \left( \frac{\omega_{kr}}{\omega_{de}} \right)^{3/2} \exp\left(-\frac{3}{4} \frac{\omega_{kr}}{\omega_{de}}\right)$, $D_1 = \frac{k_\parallel v_{the}}{\omega_{kr}}$, $D_2 = \frac{\gamma_k k_\parallel v_{the}}{\omega_{kr}^2} = \frac{k_\parallel v_{the}}{\omega_{kr}} \hat{D}_2$, and $D_3 = \frac{3\sqrt{3}\pi}{8} \frac{k_\parallel v_{the}}{|\omega_{de}|} \left( \frac{\omega_{kr}}{\omega_{de}} \right)^{3/2} \exp\left(-\frac{3}{4} \frac{\omega_{kr}}{\omega_{de}}\right) = \frac{k_\parallel v_{the}}{\omega_{kr}} \hat{D}_3$. The terms related to $D_1$, $D_2$, and $D_3$ are proportion to $k_\parallel$. Thus, we can easily obtain

$$\mathrm{Re}\langle \delta \hat{A}_{\parallel,k} \delta \hat{\phi}_{-k} \rangle = \frac{k_\parallel v_{the}}{\omega_{kr}} \frac{(C_1-C_2)-(\hat{D}_2+\hat{D}_3)(C_3+C_4)}{(C_1-C_2)^2+(C_3+C_4)^2} |\delta \hat{\phi}_k|^2, \quad (A9)$$

$$\mathrm{Im}\langle \delta \hat{A}_{\parallel,k} \delta \hat{\phi}_{-k} \rangle = -\frac{k_\parallel v_{the}}{\omega_{kr}} \frac{[(\hat{D}_2+\hat{D}_3)(C_1-C_2)+(C_3+C_4)]}{(C_1-C_2)^2+(C_3+C_4)^2} |\delta \hat{\phi}_k|^2, \quad (A10)$$